\documentclass[12pt, letterpaper]{article}
\usepackage{geometry}
\usepackage{amsmath}
\usepackage{graphicx}
\usepackage{float}
\usepackage{cite}
\usepackage{subcaption}
\usepackage{amssymb}
\newgeometry{vmargin={18mm}, hmargin={18mm,18mm}}

\begin{document}
\title{Exact solutions of the inhomogeneous nonlinear Schrödinger equation through supersymmetric potentials}
\author{David J. Fern\'andez C.$^{1}\footnote{david.fernandez@cinvestav.mx}$, O. Pav\'on-Torres$^{1}\footnote{omar.pavon@cinvestav.mx}$}
\date{%
\small{$^{1}$ Physics Department, Cinvestav, POB 14-740, 07000 Mexico City, Mexico\\
}}
\maketitle

\begin{abstract}
By employing supersymmetric quantum mechanics, we present a general algorithm to construct supersymmetric partner potentials and hence derive exact stationary solutions of the inhomogeneous nonlinear Schrödinger equation (INLSE). This is possible due to the connection between the INLSE and the nonlinear Schrödinger equation (NLSE), which can be established from a treatment based on Lie point symmetries and is related with Schrödinger equation, under certain conditions. As an illustrative example, we construct exact solutions for the INLSE through a Pösch-Teller potential with a single bound state.\\ \\
\textbf{Keywords:} Inhomogeneous nonlinear Schrödinger equation; supersymmetry; Lie point symmetry. 
\end{abstract}
\section{Introduction}
The most recent discoveries in nonlinear physics display a unique view of contemporary physics that is both unsettling and fascinating, in which the concept of ``soliton" has gained definitive relevance and has spread to almost every branch of knowledge \cite{Lentz2020BreakingTW, 10.1063/10.0014579, Hu2023NovelOS, Guidry:23, Singh2024, Kopyciński_2024, Xiong2024, Zou2024, Pavón-Torres_2024, susypair1, Hamdi2024, PhysRevB.110.235402, Hu_2024, Yomba2024, Moss2024}. A soliton is typically characterized as a solitary nonlinear wave that is self-localized, stable, and long-lasting, which results from the balance between dispersion and nonlinear effects, exhibiting behavior akin to particle-like dynamics. The study of solitons began in 1895 with the mathematical formulation of one of the most fundamental equations in soliton theory, the Korteweg-de Vries (KdV) equation \cite{Miles_1981, Korteweg1895XLIOT}. By solving it numerically, in 1965 Zabusky and Kruskal discovered that the stable solutions it yields possess all the previously mentioned properties, and it was at this point that they coined the term soliton to collectively describe these solutions \cite{PhysRevLett.15.240}. Years later, a decisive breakthrough occurs when Gardner et al., develop the inverse scattering method to solve nonlinear differential equations, such as the KdV equation \cite{PhysRevLett.19.1095}. Subsequently, in 1972 Zakharov and Shabat discovered soliton solutions for the nonlinear Schrödinger equation (NLSE) using the inverse scattering method \cite{Zakharov1970ExactTO}. The NLSE describes the propagation of waves in a dispersive medium, and it provides a more general framework than the KdV equation. 
It is worth to stress that the spatial variation of the nonlinear coefficients in the NLSE, especially in weakly compressible media, can significantly influence wave propagation. This sort of equation arises in various scenarios, such as the propagation of Bessel beams in nonlinear and inhomogeneous media \cite{Zhong:13}, surface gravity waves propagating along uneven bottoms \cite{gravitywaves}, solitons and complex nonlinear wave patterns in nonlinear lattices (NLs) \cite{RevModPhys.83.247}, and matter-wave solitons in Bose-Einstein condensates \cite{KENGNE20211}. Additionally, these equations have profound implications in modern research, including analogue Hawking radiation systems \cite{CRPHYS_2024__25_S2_A2_0}. 

Since then, several generalizations have arisen, such as the inhomogeneous nonlinear Schrödinger equation (INLSE) which can be found in a wide range of scenarios. The INLSE has garnered significant attention due to its central role in developing the concept of the 'nonautonomous soliton' \cite{PhysRevLett.98.074102}. Consequently, there has been a growing interest in constructing stationary solutions of INLSE and related equations. One widely employed approach is based on modulational instability, which involves a perturbative analysis of the varying coefficients \cite{Zolotovskii2013, ARSHAD20174153, 10.1063/1.4915516}. Another method consists in using symmetry reductions of this equation to identify easier forms and integrability criteria \cite{Bai2023, 2024JOpt, Rani2024}. The Lie symmetry analysis has been applied to the INLSE \cite{PhysRevLett.98.064102, MASOODKHALIQUE20094033} and was successfully extended to more general forms of the NLSE, such as the cubic-quintic NLSE \cite{PhysRevA.76.013612}. Using Lie point symmetries, a connection between the INLSE and the NLSE through Schrödinger equation can be as well found \cite{PhysRevLett.98.064102}. This offers an ideal framework, as for a long time exact solutions were sought for the NLSE with PT-symmetric potentials, with the aim of later implementing supersymmetric quantum mechanics (SUSY QM) to generate new potentials and stationary solutions \cite{KHARE20122880, PhysRevE.92.042901, PhysRevA.92.023821, WEN20152025}. So far, its implementation had been almost exclusively applied to the NLSE with power-law nonlinearities and complex potentials as exactly solvable examples, along with their corresponding applications \cite{Cooper2017, Nath:2017obr, Dawson_2017, PhysRevE.106.024206, MANIKANDAN2022168703, e39cce5cb1304f1ca1a32315469d6b1d}. PT-symmetric potentials have the particularity of being in general complex, whose imaginary component is associated with loss and gain terms in the NLSE. This pure imaginary part of the potential can be incorporated through a more general formalism within the framework of the INLSE. 

In the present work, a general algorithm for constructing new potentials from known ones using SUSY QM will be presented, along with their corresponding solutions to the INLSE. The work is structured as follows: in Section 2 we present an overview of the Lie point symmetry method applied to the INLSE. This technique is used then to explore the connection between the INLSE and the NLSE through Schrödinger equation. In Section 3 we introduce the main algorithm, SUSY QM, which is subsequently employed to generate new potentials from known ones. As an illustrative example, we begin from the free particle to derive solutions of the INLSE through a Pösch-Teller potential with a single bound state. Finally, in Section 4 we present our concluding remarks, offering a comprehensive discussion and a general perspective of our results.

\section{Lie symmetry analysis for the INLSE}
Lie symmetry analysis, or Lie group analysis, is based on the idea that if a differential equation has a symmetry, then the solution of the equation can be transformed in a way that such symmetry is preserved. This means that if we know the symmetries of a differential equation, we can use them to transform the equation into a simpler form, which is easier to solve. Another key application is the construction of conservation laws.  This symmetry analysis was applied to the INLSE to establish its connection with the NLSE through Schrödinger equation \cite{PhysRevLett.98.064102}. This idea was suggested as a way to obtain exact solutions for the INLSE, nevertheless, it has not been exploited in full detail. One of the main motivations of our work is to take advantage of this connection, which basically offers the possibility to generate new partner potentials of the INLSE and its corresponding solutions. 

In order to implement the Lie symmetry analysis to our system, let us consider the INLSE expressed as follows
\begin{equation}
i\psi_{t}=-\psi_{xx}+g(x)|\psi|^{2}\psi+V_{n}(x)\psi, \label{INLSE1}
\end{equation}
where $V_{n}(x)$ is an external potential and $g(x)$ describes the spatial modulation of the nonlinearity. In order to find stationary solutions of Eq. (\ref{INLSE1}) the customary ansatz $\psi=\phi e^{-i \lambda_{n} t}$ is taken, yielding
\begin{equation}
-\phi_{xx}+V_{n}(x)\phi+g(x)\phi^{3}=\lambda_{n} \phi, \qquad \phi (\pm \infty)=0. \label{INLSE2}
\end{equation}
The introduction of the sub-index in Eqs. (\ref{INLSE1}-\ref{INLSE2}) is made for convenience, as it will be clear later when we implement SUSY QM. We assume the following Lie group of point transformations to derive symmetries of the system
\begin{subequations}
\begin{equation}
x \to x+\epsilon \xi (x, \phi)+O(\epsilon ^{2}),
\end{equation}
\begin{equation}
\phi\to\phi+\epsilon \eta (x, \phi)+O(\epsilon ^{2}), 
\end{equation}
\end{subequations}
where $\epsilon \ll 1$ is a group parameter, and $\xi$ and $\eta$ are infinitesimals to be determined. The vector field corresponding to the transformation group is
\begin{equation}
M=\xi (x, \phi )\partial/\partial x +\eta (x, \phi)\partial /\partial \phi, \label{guc1}
\end{equation}
which is a symmetry of Eq. (\ref{INLSE2}). As such equation involves second-order derivatives, we need to prolong the vector field (\ref{guc1}) to second order, thus let us consider the invariance condition
\begin{equation}
M^{[2]}\left(\phi_{xx}+(\lambda_{n}-V_{n}(x))\phi-g(x)\phi^{3}\right)=0. \label{globo1}
\end{equation} 
For our case $M ^{[2]}=M+\eta ^{(1)}(x, \phi, \phi ')\partial /\partial \phi'+\eta ^{(2)}(x, \phi, \phi ', \phi'')\partial /\partial \phi''$, where $\eta ^{(1)}$ and $\eta ^{(2)}$ are provided by
\begin{equation}
\eta^{(k)}\left(x, \phi, \phi', \phi'', ..., \phi^{k}\right)=\dfrac{d\eta ^{(k-1)}}{dx}-\phi^{k}\dfrac{d\xi (x, \phi)}{dx}
\end{equation}
with $\eta ^{(0)}=\eta(x, \phi)$. 
Once applying the invariance condition to Eq. (\ref{globo1}), we obtain the following system of equations
\begin{subequations}
\begin{equation}
\xi_{\phi\phi}=0, \label{lei1}
\end{equation}
\begin{equation}
\eta_{\phi\phi}-2\eta_{\phi x}=0,\label{lei2}
\end{equation}
\begin{equation}
2 \eta_{x\phi}-\xi_{xx}-3f \xi_{\phi}=0, \label{lei3}
\end{equation}
\begin{equation}
\eta_{xx}-\xi f_{x}-\eta f_{\phi}+\eta_{\phi}f-2\xi_{x}f=0, \label{lei4}
\end{equation}
\end{subequations}
with $f(x, \phi)=V_{n}(x)\phi+g(x)\phi^{3}-\lambda_{n}\phi$. Solving the system of equations (\ref{lei1}-\ref{lei4}), we conclude that the only Lie point symmetries of Eq. (\ref{INLSE2}) take the form
\begin{equation}
M=b(x)\dfrac{\partial}{\partial x}+c(x)\phi\dfrac{\partial}{\partial \phi} \label{hop1}
\end{equation}
with
\begin{subequations}
\begin{equation}
g(x)=\dfrac{{g}_{0}}{b(x)^{3}}\exp \left[-2C \int_{0}^{x}\dfrac{1}{b(s)}ds\right],\label{lei5}
\end{equation}
\begin{equation}
c(x)=\dfrac{1}{2}b'(x)+C,\label{lei6}
\end{equation}
and
\begin{equation}
0=c''(x)-b(x)V_{n}'(x)-2b'(x)\left[V_{n}(x)-\lambda_{n}\right],\label{lei7}
\end{equation}
\end{subequations}
for any constant $C$. Equations (\ref{lei5}-\ref{lei7}) allow us to construct pairs $\{V_{n}(x), g(x)\}$ for which a Lie point symmetry exists. This process is facilitated by performing a transformation to a new coordinate system, as suggested by Lie theory. In our case, motivated by the fact that a translational generator corresponds to a conserved quantity \cite{Leach1981}, we can define the canonical transformation as
\begin{equation}
X=f(x), \quad \quad U=n(x)\phi, \label{hop2}
\end{equation}
where $f(x)$ and $n(x)$ will be determined by requiring that an energy-type conservation law $M=\partial/\partial X$ exists in the canonical variables instead of the original generator (\ref{hop1}). Thus, combining Eq. (\ref{hop1}) with Eqs. (\ref{hop2}), we obtain
\begin{subequations}
\begin{equation}
f(x)=\int_{0}^{x}\dfrac{1}{b(s)}ds, \label{dada1}
\end{equation}
\begin{equation}
n(x)=\dfrac{1}{\sqrt{b(x)}}\exp\left[-C\int_{0}^{x}\dfrac{1}{b(s)}ds\right]. \label{dada2}
\end{equation}
\end{subequations}
It is easy to see that the choice $C=0$ makes the transformations to preserve the Hamiltonian structure, and we can rewrite Eq. (\ref{INLSE2}) in terms of the new variables $U$ and $X$, defined as $U=b^{-1/2}(x)\phi$ and $X=\int_{0}^{x}\left[1/b(s)\right]ds$. Thus, Eq. (\ref{INLSE2}) turns out to be the nonlinear Schrödinger equation in its standard form
\begin{equation}
-\dfrac{d^{2}U}{dX^{2}}+g_{0}U^{3}=EU,\label{NLSE1}
\end{equation}  
with $E$ being a constant defined as $E=[\lambda_{n}-V_{n}(x)]b(x)^{2}-\dfrac{1}{4}b'(x)^{2}+\dfrac{1}{2}b(x)b''(x)$. As it has already been shown, the Lie point symmetry allows to find a connection between the INLSE (\ref{INLSE1}) with spatially inhomogeneous nonlinearity and external potential $V_{n}(x)$, with the NLSE (\ref{NLSE1}). This allows to construct exact solutions of our initial equation (\ref{INLSE1}) by means of known solutions of the NLSE, which has been exhaustively studied in the literature. An additional connection between the NLSE and INLSE by means of Schrödinger equation can be done by setting $C=0$ in Eqs. (\ref{lei5}) and (\ref{lei6}); combining the resulting equations we obtain
\begin{equation}
g(x)=\dfrac{g_{0}}{b(x)^{3}}, \label{co1}
\end{equation}
with the relation between $b(x)$ and $V(x)$ being given by:
\begin{equation}
b'''(x)-2b(x)V_{n}'(x)+4b'(x)\lambda_{n}-4b'(x)V_{n}(x)=0.
\end{equation}
We can see that when $\rho(x)=b^{1/2}(x)$, the Ermakov-Pinney equation is obtained
\begin{equation}
\rho_{xx}+\left[\lambda_{n}-V_{n}(x)\right]\rho=E/\rho^{3},
\end{equation}
whose solutions can be constructed in the way
\begin{equation}
\rho=(\alpha \varphi_{1}^{2}+2\beta \varphi_{1}\varphi_{2}+\gamma \varphi_{2}^{2})^{1/2}, \label{i1}
\end{equation}
where $\alpha$, $\beta$, $\gamma$ are arbitrary constants, and  $\varphi_{1}$ and $\varphi_{2}$ are two linearly independent solutions of the stationary Schrödinger equation
\begin{equation}
-\varphi_{xx}+V_{n}(x)\varphi=\lambda_{n}\varphi.
\end{equation} 
This provides $E$ in Eq. (\ref{NLSE1}) as follows
\begin{equation}
E=\Gamma \Lambda ^{2}, \label{en}
\end{equation} 
with 
\begin{subequations}
\begin{equation}
\Gamma=\alpha \gamma-\beta^{2},
\end{equation}
and $\Lambda$ being the Wronskian of $\varphi_{1}$ and $\varphi_{2}$
\begin{equation}
\Lambda=W(\varphi_{1}, \varphi_{2})=\varphi_{1}\varphi_{2}'-\varphi_{2}\varphi_{1}'. \label{lk}
\end{equation}
\end{subequations}
Since both solutions $\varphi_{1}$ and $\varphi_{2}$ are linearly independent, we are able to  define a canonical transformation, consequently we can evaluate $f(x)$ in a closed form. 
To prove this assertion, we start with Eq.(\ref{dada1}) and replace the explicit form of $b(x)$, supplied by Eq. (\ref{i1}), which can be expressed as
\begin{equation}
b(x)=\gamma \varphi_{1}^{2}\left[\Theta ^{2}+\left(\dfrac{\varphi_{2}}{\varphi_{1}}+\dfrac{\beta}{\gamma}\right)^{2}\right], \label{dulce1}
\end{equation}
with
\begin{equation}
\Theta=\sqrt{\dfrac{\alpha}{\gamma}-\dfrac{\beta^{2}}{\gamma^{2}}}.\label{dulce2}
\end{equation} 
Consequently, we have
\begin{equation}
f(x)=\int_{0}^{x}\dfrac{ds}{\gamma \varphi_{1}^{2}\left[\Theta ^{2}+\left(\dfrac{\varphi_{2}}{\varphi_{1}}+\dfrac{\beta}{\gamma}\right)^{2}\right]}.
\end{equation}
With the change of variables
\begin{equation}
u=\dfrac{\varphi_{2}}{\varphi_{1}}, \qquad du=\dfrac{d}{ds}\left(\dfrac{\varphi_{2}}{\varphi_{1}}\right)ds=\dfrac{\Lambda}{\varphi_{1}^{2}}ds,
\end{equation}
where $\Lambda$ is given by Eq. (\ref{lk}), in turn out that the explicit form of $f(x)$ is obtained,
\begin{equation}
f(x)=\dfrac{1}{\Lambda \sqrt{\alpha \gamma-\beta^{2}}}\tan^{-1}\left[\dfrac{\gamma}{\sqrt{\alpha\gamma-\beta^{2}}}\left(\dfrac{\varphi_{2}}{\varphi_{1}}+\dfrac{\beta}{\gamma}\right)\right].
\end{equation}
\section{INLSE with supersymmetric potentials}
Before proceeding with our analysis, we will present a brief overview of supersymmetric quantum mechanics, which will later be applied to the INLSE. We also refer to the classic references \cite{COOPER1995267, BogdanMielnik_2004, 10.1063/1.1853203, 2010AIPC.1287....3D, C.2019} (and references therein) for a more detailed discussion of the algorithm, its motivations, and the underlying philosophy. 
\subsection{Supersymmetric Quantum Mechanics}

Let us consider the following Hamiltonians $H_{0}$ and $H_{1}$, expressed as
\begin{equation}
H_{0}=-\dfrac{d^{2}}{dx^{2}}+V_{0}, \qquad H_{1}=-\dfrac{d^{2}}{dx^{2}}+V_{1},\label{susy1}
\end{equation} 
which are intertwined by two first order differential operators as follows
\begin{equation}
H_{1}A_{1}^{+}=A_{1}^{+}H_{0}, \qquad H_{0}A_{1}^{-}=A_{1}^{-}H_{1},\label{susy2} 
\end{equation}
with 
\begin{equation}
A_{1}^{-}=\dfrac{d}{dx}+\alpha_{1}(x), \qquad A_{1}^{+}=-\dfrac{d}{dx}+\alpha_{1}(x), \label{susy3}
\end{equation}
where, if $\alpha_{1}(x)$ is a real function of $x$, then $A_{1}^{+}$ stands for the Hermitian conjugate of $A_{1}^{-}$ and vice versa. Once we replace Eqs. (\ref{susy1}) and (\ref{susy3}) in (\ref{susy2}), we obtain the relation between the potential $V_{0}$ and its supersymmetric partner $V_{1}$:
\begin{equation}
V_{1}(x)=V_{0}(x)-2\alpha'_{1}(x). \label{susy4}
\end{equation}
Moreover, the following equation is obtained
\begin{equation}
V_{1}(x)\alpha_{1}(x)-\alpha_{1}''(x)=-V_{0}'(x)+\alpha_{1}(x)V_{0}(x).\label{susy5}
\end{equation}
Replacing expression (\ref{susy4}) in Eq. (\ref{susy5}) yields 
\begin{equation}
\dfrac{d}{dx}\alpha_{1}(x) ^{2}+\alpha_{1}''(x)=V_{0}'(x).\label{susy6}
\end{equation} 
Integrating this expression, we obtain
\begin{equation}
\alpha_{1}(x) ^{2}+\alpha_{1}'(x)=V_{0}(x)-\epsilon,\label{susy7}
\end{equation}
which is a Ricatti equation. If we replace
\begin{equation}
\alpha_{1}(x)=\dfrac{u_{0}'(x)}{u_{0}(x)}=\dfrac{d}{dx}\ln u_{0}(x)\label{susy8}
\end{equation}
into the Ricatti equation (\ref{susy7}) we obtain
\begin{equation}
-u_{0}''(x)+V_{0}(x)u_{0}(x)=\varepsilon u_{0}(x),\label{mi1}
\end{equation} 
which is the initial stationary Schrödinger equation for $u_{0}(x)$ associated to the factorization energy $\epsilon$. Taking into account Eq. (\ref{susy8}), Eq. (\ref{susy4}) can be expressed as 
\begin{equation}
V_{1}(x)=V_{0}(x)-2\dfrac{d ^{2}}{dx^{2}}\ln u_{0}(x). \label{susy9}
\end{equation}
The connection with the factorization method arises from the fact that
\begin{equation}
H_{0}=A_{1}^{-}A_{1}^{+}+\varepsilon, \qquad H_{1}=A_{1}^{+}A_{1}^{-}+\varepsilon. \label{mi2}
\end{equation}
Now, let us suppose that $H_{0}$ is a given solvable Hamiltonian, with known eigenfunctions $\psi_{n}^{(0)}$ and eigenvalues $E_{n}$ such that 
\begin{equation}
H_{0}\psi_{n}^{(0)}=E_{n}\psi_{n}^{(0)}, \qquad n=0, 1, ... \label{sus1}
\end{equation}
In order to implement the intertwining proccess we take a nodeless mathematical eigenfunction $u_{0}$ associated to $\varepsilon \leq E_{0}$ (see Eq. (\ref{mi1})). Equations (\ref{susy2}), together with the factorizations (\ref{mi2}) indicate that, if $A_{1}^{+}\psi_{n}^{(0)}\neq 0$, then $\{\psi_{n} ^{(1)}=A_{1} ^{+}\psi^{(0)}_{n}/\sqrt{E_{n}-\varepsilon}, n=0, 1, 2,.. \}$ is an orthonormal set of eigenfunctions of $H_{1}$ with eigenvalues $\{E_{n}\}$. This is a basis set if there is not a square-integrable function $\psi_{\epsilon}^{(1)}$ which is orthogonal to such set. Hence, let us look for a $\psi_{\epsilon}^{(1)}$ such that
\begin{equation}
0=(\psi_{\epsilon} ^{(1)}, \psi_{n} ^{(1)}) \propto (\psi_{\epsilon} ^{(1)}, A_{1}^{+}\psi_{n} ^{(0)})=(A_{1}^{-}\psi_{\epsilon} ^{(1)}, \psi_{n} ^{(0)}).
\end{equation}
Consequently, we must have,
\begin{equation}
A_{1}^{-}\psi_{\epsilon} ^{(1)}=0, 
\end{equation}
leading to 
\begin{equation}
\psi_{\epsilon} ^{(1)}\propto e^{-\int_{0}^{x}\alpha_{1}(y)dy}=\dfrac{1}{u ^{(0)}}.
\end{equation}
Since $H_{1}\psi_{\epsilon} ^{(1)}=\varepsilon \psi_{\epsilon} ^{(1)}$, the spectrum of $H_{1}$ will depend of the normalizability of $\psi_{\epsilon} ^{(1)}$, thus several cases will be obtained. We will address all these possibilities in the following subsection. 
\subsection{An exactly solvable example}
In order to implement the general algorithm of supersymmetric quantum mechanics, we start with the free particle potential $V_{0}(x)=0$. As a first step we need to build its supersymmetric partner potential, thus we need to choose carefully the seed solution. If we replace the free particle potential $V_{0}(x)=0$ in the stationary Schrödinger equation $H_{0}u_{0}=\lambda_{0} u_{0}$ we obtain 
\begin{equation}
-\dfrac{d ^{2}}{dx^{2}}u_{0}(x)=\lambda_{0} u_{0}(x).
\end{equation} 
It is well know that $\lambda_{0}$ can take several values leading to fundamentally different seed solutions $u_{0}$, namely,
\begin{equation}
u_{0}(x)= \left\{ \begin{array}{lcc}1, x & \text{if} & \lambda_{0}=0 \\ \\   \sinh(k_{0}x), \cosh(k_{0}x) & \text{if} & \lambda_{0}=-k_{0}^{2} \\ \\ \sin (k_{0}x), \cos(k_{0}x) & \text{if} & \lambda_{0}=k_{0}^{2}    \end{array} \right. \label{al1}
\end{equation}
where $k_{0}$ is a real constant. Among the several candidates to seed solution, the chosen one has to be a nodeless function in order to avoid singularities in the built potential $V_{1}(x)$.  The only solution in Eq. (\ref{al1}) fulfilling this requirement is $u_{0}=\cosh(k_{0}x)$. Thus, after substituting such seed solution in Eq. (\ref{susy9}), together with the free particle initial potential, the supersymmetric partner potential turns out to be
\begin{equation}
V_{1}(x)=-2 \, k_{0}^{2}\, \text{sech}^{2}(k_{0}x), \label{bel1}
\end{equation}
which is the Pösch-Teller potential with a single bound state. From now on this will be the potential taken for the INLSE (\ref{INLSE1}). According to Eq. (\ref{susy1}), the corresponding stationary Schrödinger equation will be 
\begin{equation}
\left[-\dfrac{d^{2}}{dx^{2}}-2\, k_{0}^{2}\, \text{sech}^{2}(k_{0}x)\right]\psi_{1} ^{(1)}=\lambda_{1}\psi_{1}^{(1)}. \label{al2}
\end{equation}
Two different kind of solutions can be identified, according to either $\lambda_{1}=\lambda_{0}$ or $\lambda_{1}\neq \lambda_{0}$.
\subsubsection{Case with $\lambda_{1}=\lambda_{0}$}

Since $\lambda_{1}=\lambda_{0}$, Eq. (\ref{al2}) becomes:
\begin{equation}
\left[-\dfrac{d^{2}}{dx^{2}}-2k_{0}^{2}\text{sech}^{2}(k_{0}x)\right]\psi_{1} ^{(1)}=\lambda_{0}\psi_{1}^{(1)}.
\end{equation}
According to SUSY QM, one solution is given by
\begin{equation}
\psi_{\varepsilon_{1}}^{(1)}(x)=\dfrac{1}{u_{0}}=\text{sech}(k_{0}x).
\end{equation}
In order to find the second linearly independent solution required to solve the INLSE, we make use of the variation of parameters method to obtain 
\begin{equation}
\psi_{\varepsilon_{2}} ^{(1)}(x)=\psi_{\varepsilon_{1}}^{(1)}\int \dfrac{dx}{\psi^{2}_{\varepsilon_{1}^{(1)}}}=\text{sech}(k_{0}x)\left[\dfrac{\sinh(2k_{0}x)}{k_{0}}+\dfrac{1}{2}x\right],
\end{equation}
where the integration constant was chosen as zero, by convenience. Thus, our two linearly independent solutions turn out to be
\begin{subequations}
\begin{equation}
\varphi_{1}(x)=\text{sech}(k_{0}x),
\end{equation}
and 
\begin{equation}
\varphi_{2}(x)=\text{sech}(k_{0}x)\left[\dfrac{\sinh(2k_{0}x)}{k_{0}}+\dfrac{1}{2}x\right].
\end{equation}
\end{subequations}
In addition, we can see that the Wronskian becomes $W(\varphi_{1}, \varphi_{2})=\Lambda=1$, as it was expected. Replacing $\varphi_{1}$ and $\varphi_{2}$ in Eqs. (\ref{dulce1}-\ref{dulce2}), we will have
\begin{equation}
b(x)= \gamma \text{sech}^{2}(k_{0}x)\left[\Theta^{2}+\left(\dfrac{\sinh(2k_{0}x)}{k_{0}}+\dfrac{1}{2}x+\dfrac{\beta}{\gamma}\right) ^{2}\right], 
\end{equation}
from which the spatially inhomogeneous nonlinearity $g(x)$ of the INLSE with the Pösch-Teller potential (\ref{bel1}) generated through SUSY QM is obtained from Eq. (\ref{co1}), and its depicted in Fig. 1a. From the preceding form of $b(x)$ we can obtain as well the expressions for $f(x)$ and $n(x)$ corresponding to the canonical transformation, see Eqs. (\ref{dada1}) and (\ref{dada2}), leading to  
\begin{equation}
X=f(x)=\dfrac{1}{\sqrt{\alpha \gamma-\beta^{2}}}\tan^{-1}\left[\dfrac{\gamma}{\sqrt{\alpha\gamma-\beta^{2}}}\left(\dfrac{\sinh(2k_{0}x)}{k_{0}}+\dfrac{1}{2}x\right)\right],
\end{equation}
\begin{equation}
n(x)=\dfrac{1}{\sqrt{\gamma \text{sech}^{2}(k_{0}x)\left[\Theta^{2}+\left(\dfrac{\sinh(2k_{0}x)}{k_{0}}+\dfrac{1}{2}x+\dfrac{\beta}{\gamma}\right) ^{2}\right]}}.
\end{equation}
Before constructing the general solutions of the INLSE, we need first to find the solutions $U(X)$ from the standard NLSE associated to Eq. (\ref{INLSE1}). It has already been shown that, in the absence of an external potential, the INLSE can support bound states with any number of solitons, such as bright and dark solitons, Peregrine solitons, elliptic solitons, compactons, kinks, breathers, among others \cite{Hasegawa1990, Vyas2006, Akhmediev2009, Ohta2012, Raju2014, Fan2023, Pavon-Torres2024}. However, we must be careful in the opposite direction, since an infinite number of soliton solutions are not obtainable. Thus, once we have implemented SUSY QM to the free particle, the auxiliary system is now described by a Pösch-Teller potential with a single bound state. Note that, for the particular case of a hyperbolic secant potential barrier, bound states with an arbitrary number of solitons have been obtained for the inhomogeneous cubic-quintic NLSE, as well as in the case without external potential.  

Based on the above result, we generate the external potential and nontrivial nonlinearities through SUSY QM. We can see that for any given potential not all soliton solution can satisfy the boundary condition $\phi(x)\to 0$ as $x \to \pm \infty$. As can be observed in Figs. 1a, 1b, 1c, and 1d, the first three nonlinearities have a deformed sech-type of potential (see figures 1a to 1c), and the last potential has a quasi-periodic behaviour (compare Fig. 1d). Thus, we need to consider as a soliton solution the following one

\begin{equation}
U(X)=\eta \dfrac{\text{sn}(\mu X, k)}{\text{dn}(\mu X, k)} \label{luah00}
\end{equation}
where $\text{sn}(\mu X, k)$ and $\text{dn}(\mu X, k)$ are Jacobi elliptic functions,
\begin{equation}
\mu^{2}=\dfrac{E}{1-2k ^{2}}, \quad \quad \eta^{2}=\dfrac{2k^{2}(k^{2}-1)}{|g_{0}|}\mu^{2}, \label{luah0}
\end{equation}
and $E$ is given by Eq. (\ref{en}). As we need to lie in the bounded domain of the NLSE, $X$ must fulfil
\begin{equation}
-\xi_{1}\leq X \leq \xi_{1}, \label{luah1}
\end{equation}
with $\xi_{1}$ being given by
\begin{equation}
\xi_{1}=\dfrac{\pi}{2}\cdot\dfrac{1}{\Lambda \sqrt{\alpha\gamma-\beta ^{2}}}.
\end{equation}
We have to make $U(\pm \xi_{1})=0$ by assuming that $\mu \xi_{1}=2n K(k)$ $(n=1, 2, ...)$ for satisfying the boundary condition $\phi (\pm \infty)=0$, where $\mu$ is given by Eq. (\ref{luah0}), $K(k)$ stands for the elliptic integral, 
\begin{equation}
K(k)=\int_{0}^{\pi/2}\dfrac{1}{\sqrt{1-k^{2}\sin^{2} (x)}}dx,
\end{equation}
and $k$ is the elliptic modulus. Thus, the explicit expression for the modified soliton becomes
\begin{equation}
\phi(x)=\eta\sqrt{\gamma \text{sech}^{2}(k_{0}x)\left[\Theta^{2}+\left(\dfrac{\sinh(2k_{0}x)}{k_{0}}+\dfrac{1}{2}x+\dfrac{\beta}{\gamma}\right) ^{2}\right]}\dfrac{\text{sn}(\mu X, k)}{\text{dn}(\mu X, k)},
\end{equation}
which is schematically depicted in Fig. 1b. 
\begin{figure}[h]
    \centering
    \begin{subfigure}[b]{0.5\textwidth}
        \centering
        \includegraphics[height=2.2in]{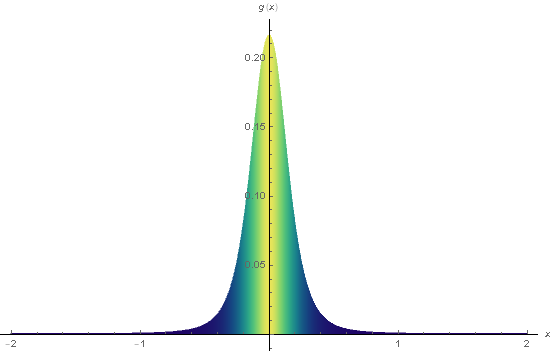}
        \caption{}\label{fig:sn1}
    \end{subfigure}%
    ~ 
    \begin{subfigure}[b]{0.5\textwidth}
        \centering
        \includegraphics[height=2.2in]{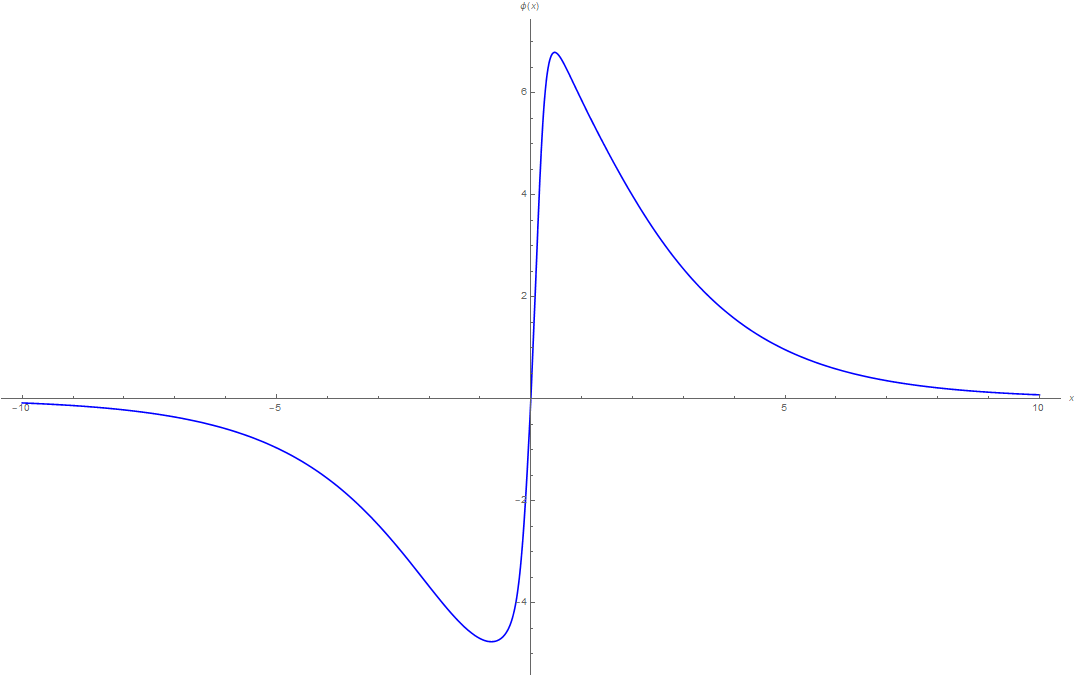}
        \caption{}\label{fig:sn2}
    \end{subfigure}
    \caption{(a) Spatially inhomogeneous nonlinearity $g(x)$. (b) Soliton profile for the Pösch-Teller potential with one bound state. The chosen parameters are $k_{0}=0.5$, $\alpha=2$, $\gamma=3$ and $\beta=g_{0}=1$.}
\end{figure}

\subsubsection{Case with $\lambda_{1}\neq\lambda_{0}$}
According to SUSY QM with $\lambda_{1}\neq\lambda_{0}$, we can choose as the two linearly independent solutions $\varphi_{1}$ and $\varphi_{2}$ the result of acting the intertwining operator $A_{1} ^{+}$ onto each of the two linearly independent solutions reported in each row of Eq. (\ref{al1}), with the change $\lambda_{0}\to \lambda_{1}$. This leads to three different cases.\\
\\
\textbf{Case 2a.} For $\lambda_{1}=0$, we will get\\
\begin{subequations}
\begin{equation}
\varphi_{1}(x)=k_{0} \tanh (k_{0}x),
\end{equation}
\begin{equation}
\varphi_{2}(x)=-1+k_{0}\, x \tanh (k_{0}x),
\end{equation}
\end{subequations}
as the two linearly independent solutions, with $\Lambda=W(\varphi_{1}, \varphi_{2})=k_{0}^{2}$. Once we replace both solutions in Eq. (\ref{dulce1}), it is obtained\\
\begin{equation}
b(x)= \gamma \text{tanh}^{2}(k_{0}x)\left[\Theta^{2}+\left(-\dfrac{1}{k_{0}}\text{coth}(k_{0}x)+x+\dfrac{\beta}{\gamma}\right) ^{2}\right].
\end{equation}
The spatially inhomogeneous nonlinearity $g(x)$ for the INLSE with the Pösch-Teller potential can be obtained from Eq. (\ref{co1}), and it is depicted in Fig. 2a. As in our previous example, we obtain $b(x)$, and consequently $f(x)$ and $n(x)$, from Eqs. (\ref{dada1}) and (\ref{dada2}): 
\begin{equation}
X=f(x)=\dfrac{1}{k_{0}^{2}\sqrt{\alpha \gamma-\beta^{2}}}\tan^{-1}\left[\dfrac{\gamma}{\sqrt{\alpha\gamma-\beta^{2}}}\left(-\dfrac{1}{k_{0}}\text{coth}(k_{0}x)+x+\dfrac{\beta}{\gamma}\right)\right],
\end{equation}
\begin{equation}
n(x)=\dfrac{1}{\sqrt{\gamma \text{tanh}^{2}(k_{0}x)\left[\Theta^{2}+\left(-\dfrac{1}{k_{0}}\text{coth}(k_{0}x)+x+\dfrac{\beta}{\gamma}\right) ^{2}\right]}}.
\end{equation}
The corresponding soliton solution, modified by the nonlinear inhomogeneity, can be constructed in a similar way as in the preceding case, with $U(X)$ given by Eqs. (\ref{luah00}-\ref{luah1}), and (\ref{en}). Therefore, the modified soliton turns out to be 
\begin{equation}
\phi(x)=\eta \sqrt{\gamma \text{tanh}^{2}(k_{0}x)\left[\Theta^{2}+\left(-\dfrac{1}{k_{0}}\text{coth}(k_{0}x)+x+\dfrac{\beta}{\gamma}\right) ^{2}\right]} \dfrac{\text{sn}(\mu X, k )}{\text{dn}(\mu X, k)}, 
\end{equation}
which is drawn in Fig. 2b. 
\begin{figure}[h]
    \centering
    \begin{subfigure}[b]{0.5\textwidth}
        \centering
        \includegraphics[height=2.2in]{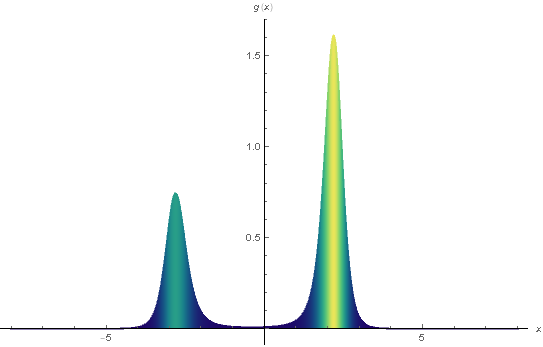}
        \caption{}\label{fig:cn1}
    \end{subfigure}%
    ~ 
    \begin{subfigure}[b]{0.5\textwidth}
        \centering
        \includegraphics[height=2.2in]{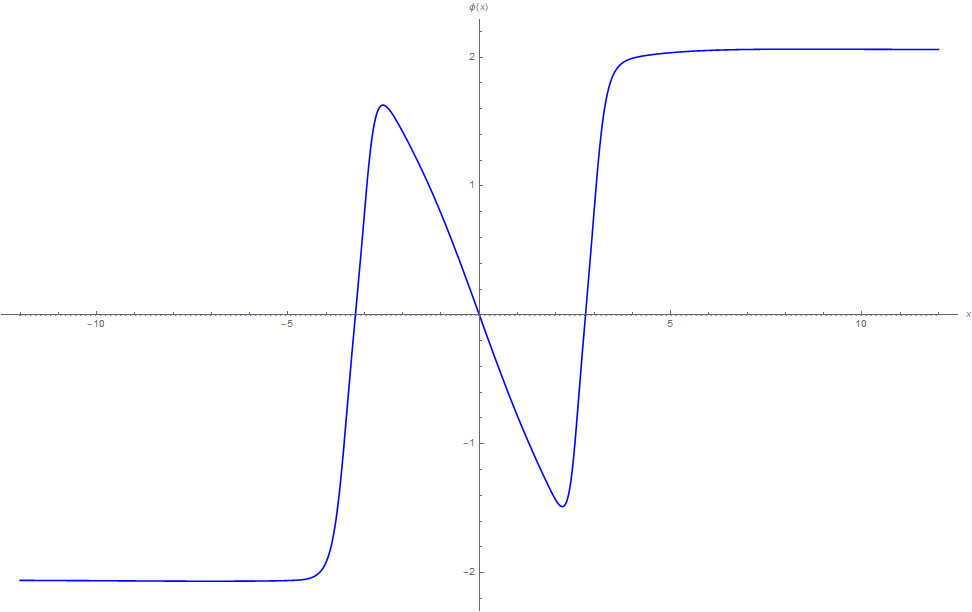}
        \caption{}\label{fig:cn2}
    \end{subfigure}
    \caption{(a) Spatially inhomogeneous nonlinearity $g(x)$. (b) Dark soliton profile for the Pösch-Teller potential with one bound state. The chosen parameters are $k_{0}=0.5$, $\alpha=2$, $\gamma=3$ and $\beta=g_{0}=1$.}
\end{figure}\\
\textbf{Case 2b.} For $\lambda_{1}=-k_{1}^{2}$ we have the two linearly independent solutions\\
\begin{subequations}
\begin{equation}
\varphi_{1}(x)=-k_{1}\cosh(k_{1}x)+k_{0}\tanh(k_{0}x)\sinh(k_{1}x),
\end{equation}
\begin{equation}
\varphi_{2}(x)=-k_{1}\sinh(k_{1}x)+k_{0}\tanh(k_{0}x)\cosh(k_{1}x),
\end{equation}
\end{subequations}
with $\Lambda=W(\varphi_{1}, \varphi_{2})=k_{1}(k_{1}^{2}-k_{0}^{2})$. Once we replace both solutions in Eq. (\ref{dulce1}), it is obtained\\
\begin{align}
b(x) &=\gamma (-k_{1}\cosh(k_{1}x)+k_{0}\tanh(k_{0}x)\sinh(k_{1}x)) ^{2} \nonumber\\
       & \quad \times \left[\Theta^{2}+\left(\dfrac{-k_{1}\sinh(k_{1}x)+k_{0}\tanh(k_{0}x)\cosh(k_{1}x)}{-k_{1}\cosh(k_{1}x)+k_{0}\tanh(k_{0}x)\sinh(k_{1}x)}+\dfrac{\beta}{\gamma}\right) ^{2}\right],
\end{align}
from which the spatially inhomogeneous nonlinearity $g(x)$ for the INLSE with the Pösch-Teller potential is obtained from Eq. (\ref{co1}), and it is depicted in Fig. 3a. Once again, from this $b(x)$ we obtain easily the canonical transformation function $f(x)$ from Eq. (\ref{dada1}), leading to
\begin{equation}
X=f(x)=\dfrac{1}{\Lambda\sqrt{\alpha \gamma-\beta^{2}}}\tan^{-1}\left[\dfrac{\gamma}{\sqrt{\alpha\gamma-\beta^{2}}}\left(\dfrac{-k_{1}\sinh(k_{1}x)+k_{0}\tanh(k_{0}x)\cosh(k_{1}x)}{-k_{1}\cosh(k_{1}x)+k_{0}\tanh(k_{0}x)\sinh(k_{1}x)}+\dfrac{\beta}{\gamma}\right)\right].
\end{equation}
As previously, $n(x)$ and $g(x)$ are obtained from expressions (\ref{dada1}) and (\ref{dada2}), respectively. In addition, the corresponding soliton solution modified by the nonlinear inhomogeneity is constructed in a similar way as in the preceding case, with $U(X)$ given by Eqs. (\ref{luah00}-\ref{luah1}) and (\ref{en}), leading to 
\begin{equation}
\phi(x)=b(x)^{1/2}\eta\dfrac{\text{sn}(\mu X, k)}{\text{dn}(\mu X, k)},
\end{equation}
which is depicted in Fig. 3b. 
\begin{figure}[h]
    \centering
    \begin{subfigure}[b]{0.5\textwidth}
        \centering
        \includegraphics[height=2.2in]{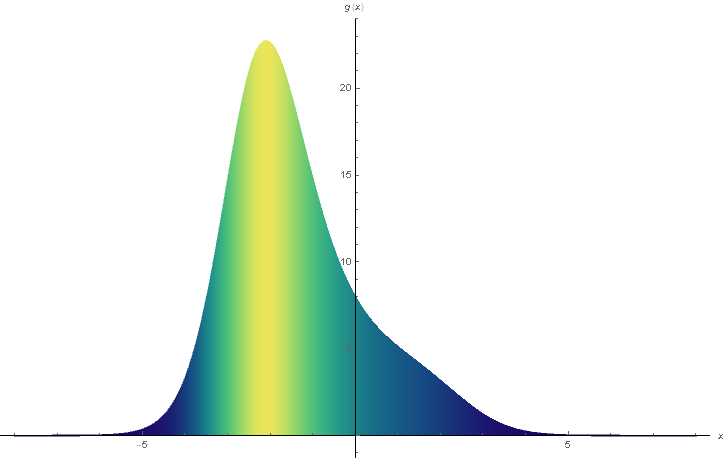}
        \caption{}\label{fig:cc1}
    \end{subfigure}%
    ~ 
    \begin{subfigure}[b]{0.5\textwidth}
        \centering
        \includegraphics[height=2.2in]{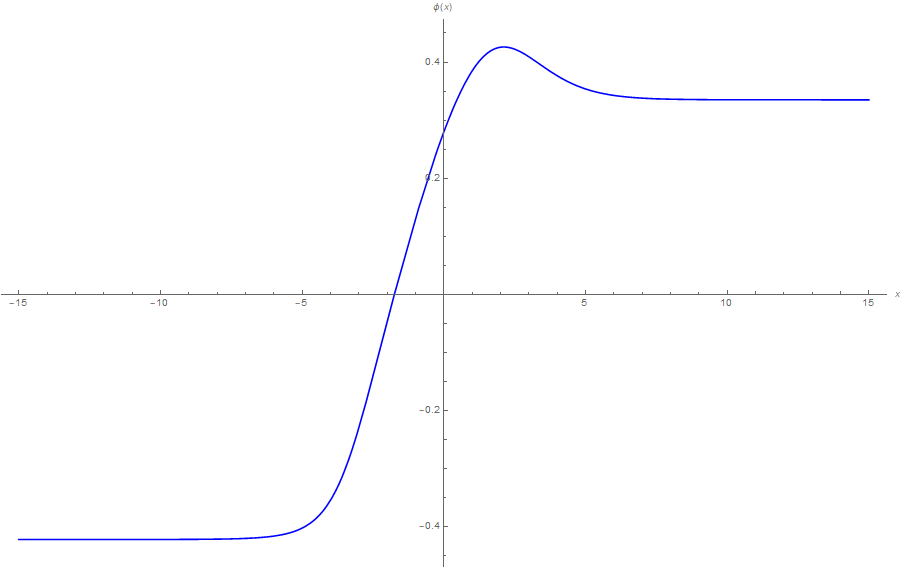}
        \caption{}\label{fig:cc2}
    \end{subfigure}
    \caption{(a) Spatially inhomogeneous nonlinearity $g(x)$. (b) Dark soliton profile for the Pösch-Teller potential with one bound state. The chosen parameters are $k_{0}=0.5$, $k_{1}=0.4$, $\alpha=2$, $\gamma=3$ and $\beta=g_{0}=1$.}
\end{figure}\\
\textbf{Case 2c.} For $\lambda_{1}=k_{1}^{2}$ we will have\\
\begin{subequations}
\begin{equation}
\varphi_{1}(x)=k_{1}\sin(k_{1}x)+k_{0}\tanh(k_{0}x)\cos(k_{1}x),
\end{equation}
\begin{equation}
\varphi_{2}(x)=-k_{1}\cos(k_{1}x)+k_{0}\tanh(k_{0}x)\sin(k_{1}x),
\end{equation}
\end{subequations}
with $\Lambda=W(\varphi_{1}, \varphi_{2})=k_{1}(k_{1}^{2}+k_{0}^{2})$. Once we replace both solutions in Eq. (\ref{dulce1}), it is obtained\\
\begin{align}
b(x) &=\gamma (k_{1}\sin(k_{1}x)+k_{0}\tanh(k_{0}x)\cos(k_{1}x)) ^{2}\nonumber\\
       & \quad \times \left[\Theta^{2}+\left(\dfrac{-k_{1}\cos(k_{1}x)+k_{0}\tanh(k_{0}x)\sin(k_{1}x)}{k_{1}\sin(k_{1}x)+k_{0}\tanh(k_{0}x)\cos(k_{1}x)}+\dfrac{\beta}{\gamma}\right) ^{2}\right],
\end{align}
from which the spatially inhomogeneous nonlinearity $g(x)$ for the INLSE with the Pösch-Teller potential is obtained from Eq. (\ref{co1}), and it is depicted in Fig. 4a. The inhomogeneous nonlinearity has a periodic behaviour as $x\to \pm \infty$, thus it is expected that the soliton behaves similarly to the case of a fully periodic inhomogeneous nonlinearity obtained from the free particle potential. However, an additional modification is expected, since we have obtained a deformed nonlinearity. As in the previous subcases, once $b(x)$ is obtained we can calculate easily $f(x)$ and $n(x)$ from Eqs. (\ref{dada1}) and (\ref{dada2}), respectively, leading to 
\begin{equation}
X=f(x)=\dfrac{1}{\Lambda\sqrt{\alpha \gamma-\beta^{2}}}\tan^{-1}\left[\dfrac{\gamma}{\sqrt{\alpha\gamma-\beta^{2}}}\left(\dfrac{-k_{1}\cos(k_{1}x)+k_{0}\tanh(k_{0}x)\sin(k_{1}x)}{k_{1}\sin(k_{1}x)+k_{0}\tanh(k_{0}x)\cos(k_{1}x)}+\dfrac{\beta}{\gamma}\right)\right]. \label{x2c}
\end{equation}
As it was already told, we have gotten a quasi-periodic inhomogeneous nonlinearity, thus we can naturally use a kink soliton profile similar to one of reference \cite{PhysRevLett.98.064102}, which is given by 
\begin{equation}
U(X)=\sqrt{\dfrac{E}{2}}\tanh\left(\sqrt{\dfrac{E}{2}}X\right),
\end{equation}
where $X$ is given by Eq. (\ref{x2c}). Therefore, we have that the soliton modified by the nonlinearity generated by supersymmetry acquires the form 
\begin{equation}
\phi(x)=b ^{1/2}(x)U(X),
\end{equation}
which is depicted schematically in Fig. 4b. The quasi-periodity of the nonlinearity generated by supersymmetry affects directly the infinite tails of the soliton profile, and the sech-type of inhomogeneity affects only the core of the dark soliton. 
\begin{figure}[h]
    \centering
    \begin{subfigure}[b]{0.5\textwidth}
        \centering
        \includegraphics[height=2.2in]{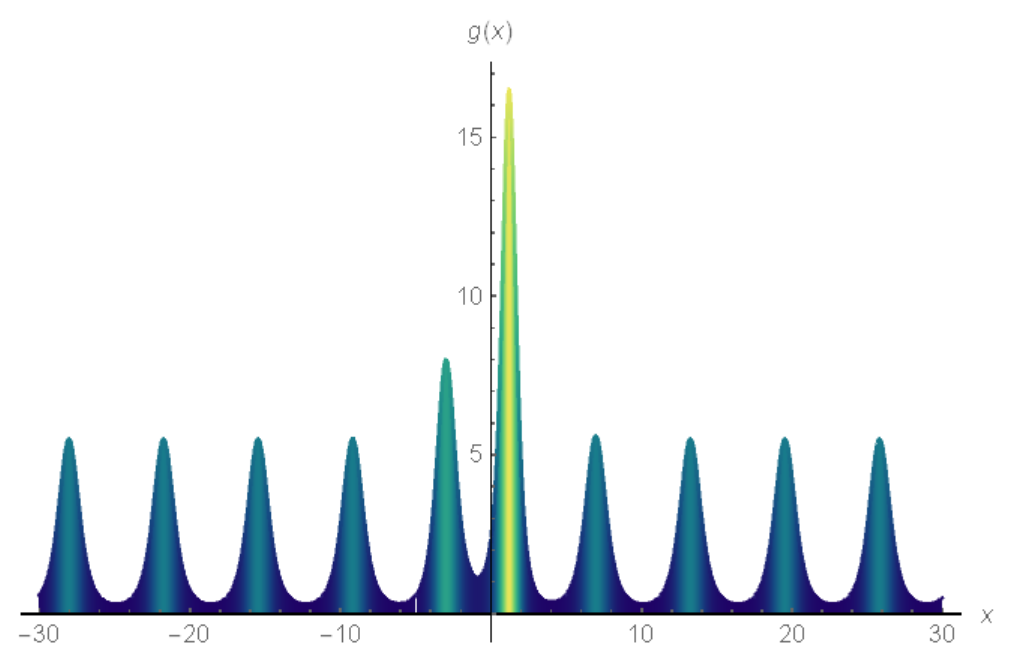}
        \caption{}\label{fig:dn1}
    \end{subfigure}%
    ~ 
    \begin{subfigure}[b]{0.5\textwidth}
        \centering
        \includegraphics[height=2.2in]{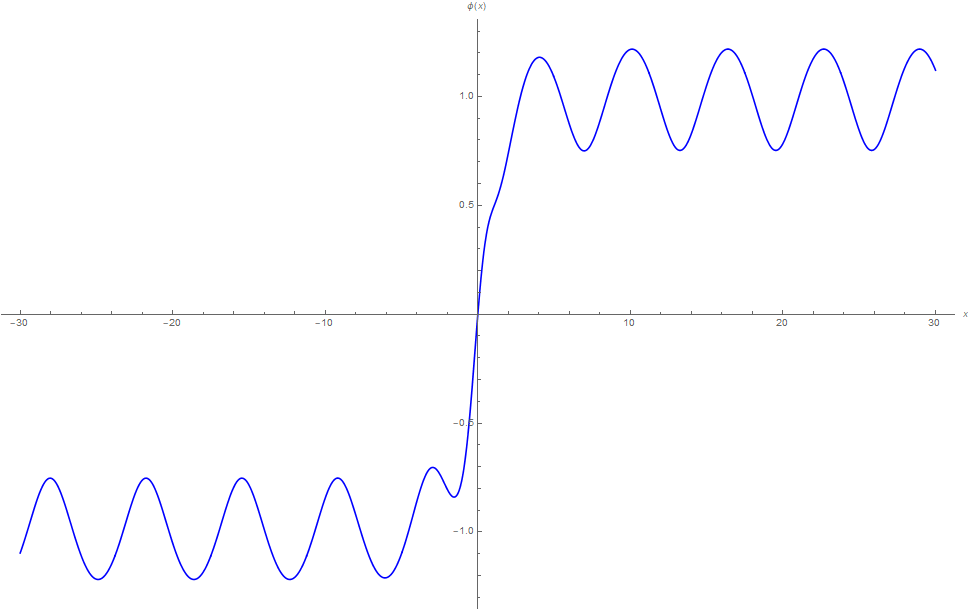}
        \caption{}\label{fig:dn2}
    \end{subfigure}
    \caption{(a) Spatially inhomogeneous nonlinearity $g(x)$. (b) Dark soliton profile for the Pösch-Teller potential with one bound state. The chosen parameters are $k_{0}=0.5$, $k_{1}=0.4$, $\alpha=2$, $\gamma=3$ and $\beta=g_{0}=1$.}
\end{figure}

\section{Conclusion}
A general algorithm has been implemented to construct exact solutions of the INLSE, based on supersymmetric quantum mechanics and the Lie symmetry analysis of such equation. The related inhomogeneous nonlinearity and the potential employed allows to use the algorithm in a especially simple way. Nevertheless, its implementation yields more intricate forms for the nonlinearity, which could represent an inconvenience, since the physical situations should determine this form. In other words, the algorithm and its implementation will depend on the physical problem under study. For example, the spatially inhomogeneous nonlinearity is mainly due to the magnetic field or the laser intensity acting on the Feschbach resonances. Thus, our spatially inhomogeneous nonlinearity can be generated by controlling the Feschbach resonances optically through different types of beams.  In our particular exactly solvable example we have considered the simplest potential at hand, the free particle, to generate its supersymmetric partner potentials, and corresponding exact solutions. Finally, although we considered here first-order transformations, we can easily generalize the scheme to higher order transformations, to build potentials with $n$ bound states departing from the free particle case. Our present study opens as well the possibility to employ not only real potentials, but also complex potentials as the one studied in the context of PT symmetry. 
\section*{Acknowledgements}
The authors acknowledge the support of Secretaria de Ciencia, Humanidades, Tecnología e Innovación (SECIHTI-México), grant FORDECYT-PRONACES/61533/2020. OPT acknowledges SECIHTI for a postdoctoral fellowship.

\section*{Data availability statement}
All data that support the findings of this study are included within the article (and any supplementary
files).
 
\section*{Ethical approval}
Not applicable

\section*{Competing interests}
The authors have no conflicts of interest to disclose.

\bibliographystyle{unsrt}
\bibliography{susy1}

\end{document}